\documentclass[12pt]{iopart}
\usepackage[figuresright]{rotating}
\begin{document}
\title{Jet Quenching in Non-Conformal Holography}
\author{Andrej Ficnar$^1$, Jorge Noronha$^2$ and Miklos Gyulassy$^1$}
\address{$^1$ Department of Physics, Columbia University, New York, NY 10027, USA}
\address{$^2$ Instituto de F\'{i}sica, Universidade de S\~{a}o Paulo, 05315-970 S\~{a}o Paulo, SP, Brazil}
\ead{aficnar@phys.columbia.edu}
\begin{abstract}
We use our non-conformal holographic bottom-up model for QCD described in \cite{our-paper} to further study the effect of the QCD trace anomaly on the energy loss of both light and heavy quarks in a strongly coupled plasma. We compute the nuclear modification factor $R_{AA}$ for bottom and charm quarks in an expanding plasma with Glauber initial conditions. We find that the maximum stopping distance of light quarks in a non-conformal plasma scales with the energy with a temperature (and energy) dependent effective power.
\end{abstract}

\submitto{\jpg}

\section{Introduction}
In \cite{our-paper} we presented a non-conformal holographic bottom-up model that captures some of the phenomenological properties of QCD at finite temperature. In this model, asymptotic freedom is replaced by conformal invariance above a certain UV scale. The main idea is to break the conformal invariance of $\mathcal{N}=4$ SYM field theory from the original AdS/CFT correspondence \cite{maldacena} in order to study effects from the QCD trace anomaly and its thermodynamical consequences as closely as possible. Our model is based on a bottom-up approach to this problem (for an alternative, top-down approach, see e.g. \cite{mohammed}) where we consider an effective, 5-dimensional gravity theory coupled to a scalar field \cite{scalarfieldmodels}. The potential for the scalar field $V(\phi)$ breaks conformal invariance in the infrared and it is engineered to holographically reproduce some thermodynamical properties of finite-temperature QCD \cite{Noronha:2009ud} that have been computed on the lattice \cite{bnl-columbia}. The space-time geometry is asymptotically $AdS_5$, which is translated into conformal invariance of the dual field theory in the UV.

\section{Heavy quark energy loss and $R_{AA}$}
A heavy quark of mass $m_Q$ is dual to a string in the bulk which stretches from the bottom of a D4-brane located at a radial coordinate $r_m\sim 1/m_Q$ (in coordinates where the boundary is at $r=0$) to the black brane horizon at $r_H$ \cite{karch-katz}. Using the trailing string ansatz, we can study the heavy quark energy loss \cite{e-loss} in our model at various temperatures and momenta, and significant deviations from the results in the conformal limit have been observed around $T_c$ \cite{our-paper}.

Here we use this setup to compute the nuclear modification factor $R_{AA}$ of charm and bottom quarks (for similar calculations in a conformal plasma see \cite{conformalRAA}). We use Glauber initial conditions where the number density of binary collisions $T_{AA}$ and the participant nucleon density $\rho_{part}$ are obtained from a Woods-Saxon nuclear distribution. A jet produced at a point $\vec{x}_\perp$ in the transverse plane and moving in the azimuthal direction $\phi$ ``sees'' at time $t$ the temperature $T(\vec{x}_\perp,\phi,t)\propto \left[\rho_{part}\left(\vec{x}_\perp+(t-t_i)\hat{e}(\phi)\right)/t\right]^{1/3}$, where $t_i$ is the initial jet production time, which we chose to be 1 fm/c. To compute $R_{AA}$ as a function of $p_{T,f}$, the final momentum of the quenched jet, one must know its corresponding initial momentum $p_{T,i}(p_{T,f},\vec{x}_\perp,\phi)$. To obtain that, one must, for a given $\vec{x}_\perp$ and $\phi$, solve the differential equation $\frac{dp(t)}{dt}=\left[\frac{dE}{dx}\right](p(t),T(\vec{x}_\perp,\phi,t))$, where $\left[\frac{dE}{dx}\right](p,T)$ is the energy loss as a (numerical) function of momentum and temperature obtained from our model. This equation is solved with the condition $p(t_f)\equiv p_{T,f}$, where the final time $t_f$ is defined by a standard freeze-out condition, $T(\vec{x}_\perp,\phi,t_f)=T_{fo}$, where we chose $T_{fo}=150$ MeV. The initial momentum is then obtained as $p_{T,i}=p(t_i)$. Finally, one must average over the azimuthal directions $\phi$ and the transverse production points $\vec{x}_\perp$:
\begin{equation}\label{eq3}
R_{AA}^Q(p_{T,f})=\int d^2\vec{x}_\perp \frac{T_{AA}(\vec{x}_\perp)}{N_{bin}}\int\limits_0^{2\pi}\frac{d\phi}{2\pi}\frac{\frac{d\sigma_Q}{dydp_T}(p_{T,i})}{\frac{d\sigma_Q}{dydp_T}(p_{T,f})}\frac{dp_{T,i}}{dp_{T,f}}
\end{equation}
where $N_{bin}$ is the total number of binary collisions and $\frac{d\sigma_Q}{dydp_T}$ are distribution functions obtained from the FONLL production cross sections \cite{fonll}. 

We computed $R_{AA}$ for heavy quarks at RHIC and LHC energies with the initial temperatures at the center of the plasma of 265 MeV and 357 MeV and nucleon-nucleon inelastic cross-sections $\sigma_{NN}=$ 4.2 and 6.3 fm$^2$ \cite{xin-nian}, respectively (Fig.\ 1, left plot). For bottom quarks, $R_{AA}$ flattens out already at $p_T \sim 10$ GeV, which is a consequence of the asymptotic conformal invariance of our model. Also, note that the values of $R_{AA}$ for RHIC and LHC are very similar. In fact, even though the maximum temperature at LHC is higher than at RHIC, the spectral indices at RHIC are steeper than at LHC and, thus, the heavy quark $R_{AA}$ does not change much when one increases $\sqrt{s}$ by one order of magnitude.

For charm quarks $R_{AA}$ is relatively small, which is simply due to the fact that the trailing string energy loss in {\it any} holographic model decreases monotonically with mass. However, one should be careful when using the trailing string ansatz in this case since already for $T\sim$ 290 MeV, $r_H\approx r_m$ for the charm quark, which implies that the trailing string ansatz is not applicable in this case. 
\begin{figure}[h!]
	  \centering
    \includegraphics[width=51mm]{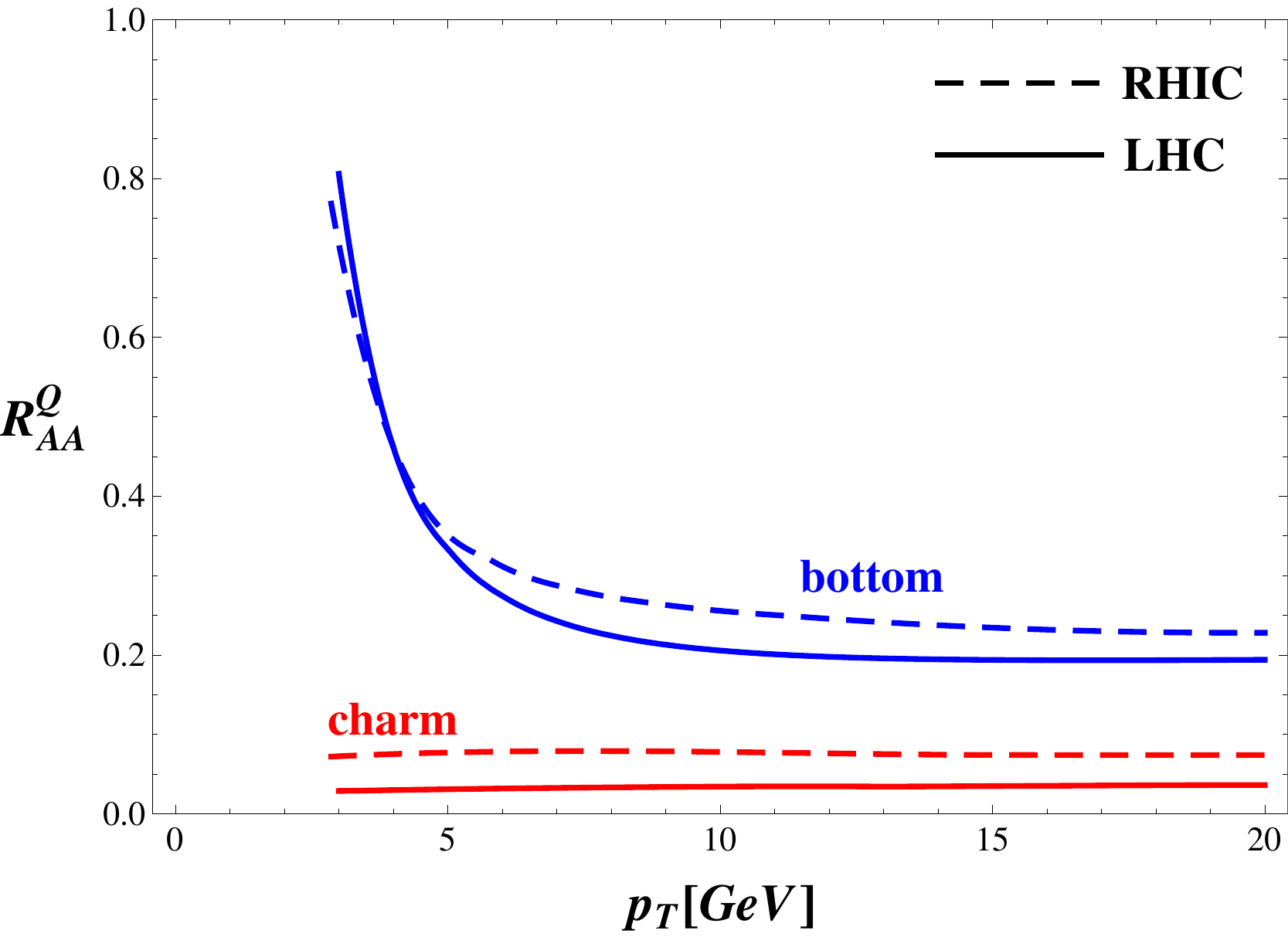}
    \includegraphics[width=51mm]{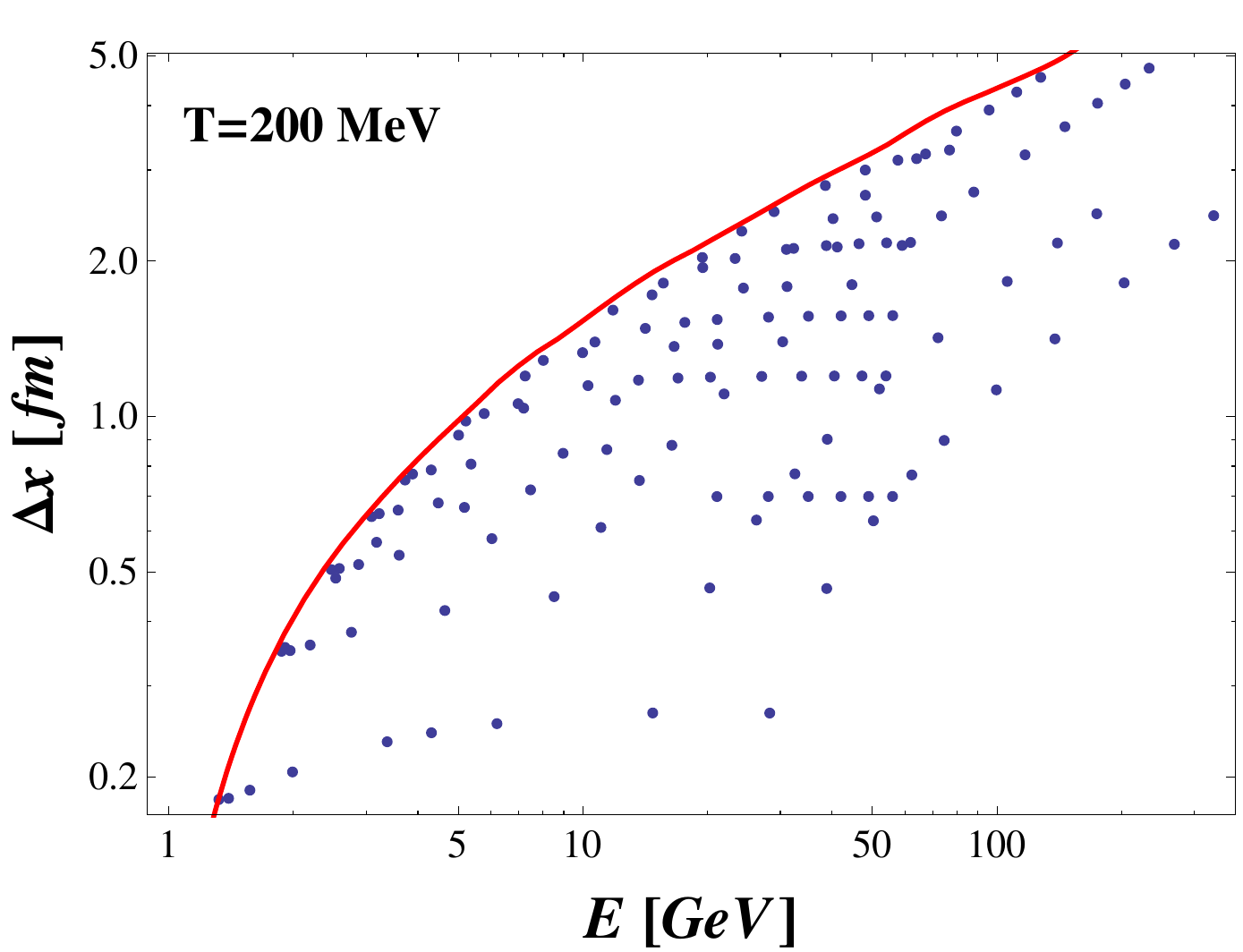}
    \includegraphics[width=51mm]{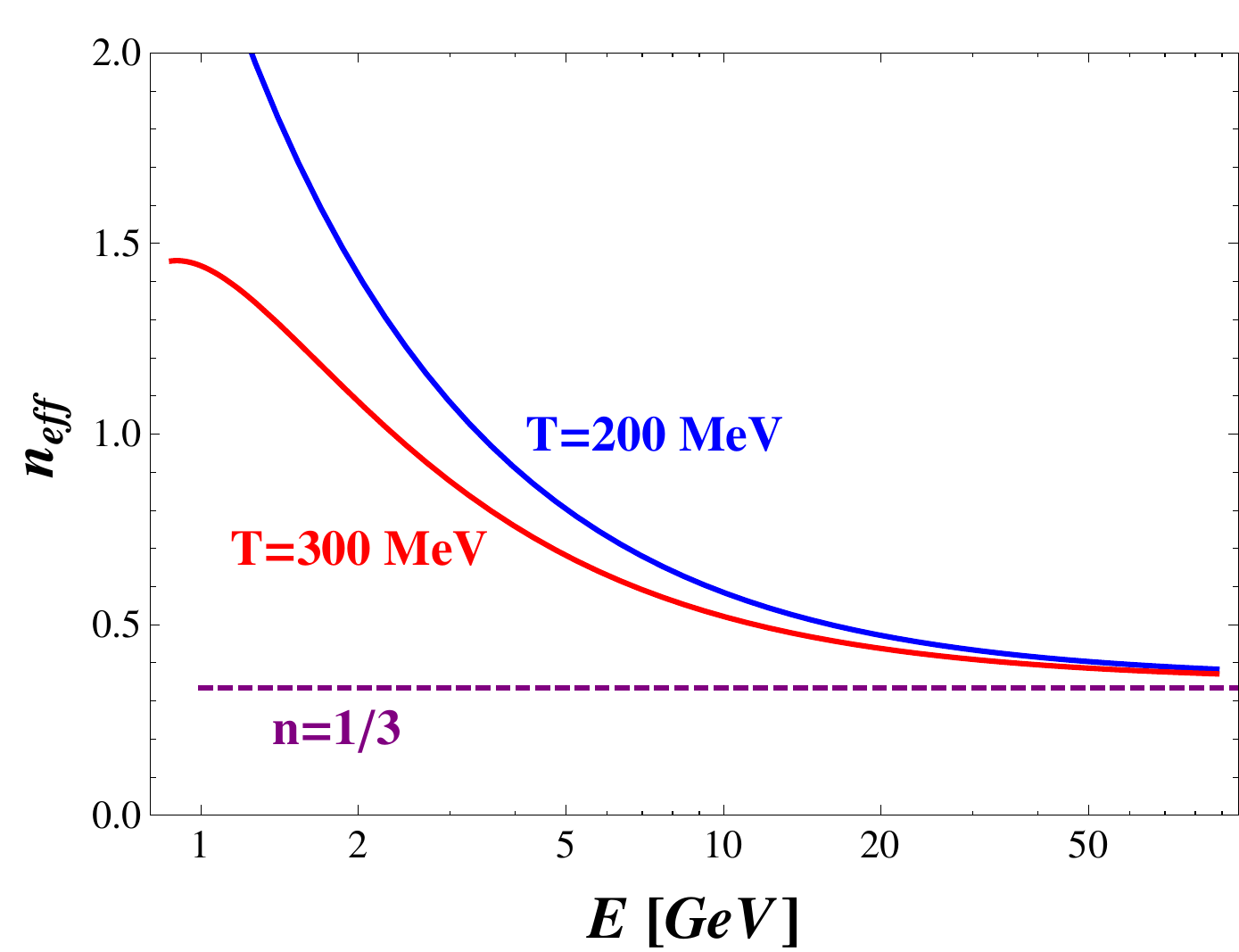}\\
    \vskip 0.1in
    \parbox[l]{162mm}{\footnotesize{\textbf{Figure 1.}} \textit{Left:} Nuclear modification factor $R_{AA}$ at impact parameter $b=3$ fm for charm and bottom quarks in an expanding plasma as a function of the transverse momentum $p_T$ for RHIC (dashed curve) and LHC (solid curve). \textit{Center:} Light quark stopping distance $\Delta x$ as a function of energy $E$ for different initial conditions (blue points) and the extracted envelope for the maximum stopping distance (red curve). \textit{Right:} Effective stopping power $n_{eff}$ for light quarks as a function of energy $E$ for two different temperatures.}
\end{figure}

\section{Light quark energy loss and the maximum stopping distance}
For light quarks, the bottom of the D4-brane $r_m\sim1/m_Q$ can approach the black brane event horizon $r_H$. In this case we follow \cite{chesler} and study open strings with both endpoints on the flavor brane, which fall freely towards the black brane. These strings are dual to dressed $q\bar{q}$ pairs and the total spatial distance the string endpoint traverses can be identified with the stopping distance of light quarks in a strongly-coupled plasma.

For a given initial energy of the string, this stopping distance will vary with the string initial conditions (initial spatial and velocity profile). However, it is still possible to obtain the maximum stopping distance for a given energy. In fact, it was shown in \cite{Gubser:2008as,chesler} that this quantity scales as $\Delta x_{max}^{AdS}\sim E^{1/3}$ in a conformal plasma. However, in our model, this power gets modified and the effective power becomes temperature and energy-dependent, 
$\Delta x_{max}\sim E^{n_{eff}(T,E)}$ (Fig.\ 1, center and right plots). We see that for large enough energies this effective power asymptotes to the conformal value of $1/3$ and also that the effective power is closer to the $1/3$ limit the further away we are from $T_c$, which are both consequences of the asymptotic conformal invariance of our model. 

Note, however, that this maximum stopping distance is not a {\it typical} stopping distance of light quarks; it is a rather crude quantity, which might be used as a
phenomenological guideline, but should not be used to obtain the instantaneous energy loss that enters in, for example, calculations of $R_{AA}$. Instantaneous energy loss, defined as the usual $\Pi_x^r$ worldsheet current, becomes a non-trivial quantity in this case since it is explicitly time-dependent and also depends on the point on the string on which one evaluates it. As shown in \cite{chesler}, at late times the instantaneous energy loss develops a Bragg-like peak but its magnitude depends on the point on the string where the energy loss is evaluated. 

Furthermore, the amount of instantaneous energy loss depends heavily on the initial conditions of the string (which are dual to the initial gauge field configuration associated with the $q\bar{q}$ pair). These two problems are the main obstacles that need to be overcome in order to obtain a consistent description of light quark energy loss that can be used in the calculation of observables such as $R_{AA}$.

\section{Conclusions, Outlook, and Acknowledgments}

In this proceedings we have extended our previous analysis of heavy quark energy loss in non-conformal plasmas started in \cite{our-paper} and computed the partonic $R_{AA}$ in an expanding plasma with Glauber initial conditions. The bottom quark $R_{AA}$ at RHIC decreases with $p_T$ and asymptotes to a value of about 0.2. Charm $R_{AA}$ is predicted to be heavily suppressed ($< 0.1$), but one should keep in mind that the trailing string model may not be applicable to describe charm quarks. According to our calculations, the heavy quark $R_{AA} \times p_T$ remains practically unaltered when going from RHIC to LHC energies.

We also showed that the typical effective power in the energy dependence of the maximum stopping distance of light quarks becomes energy and temperature dependent in a non-conformal plasma. A consistent treatment of instantaneous energy loss of light quarks has some practical difficulties and possible solutions are currently being explored \cite{our-future-paper}. 

We thank M.~Mia, A.~Buzzatti, R.~Pisarski, and P.~Chesler for helpful discussions. A.F. and M.G. acknowledge support by US-DOE Nuclear Science Grant No. DE-FG02-93ER40764.


\end{document}